\documentclass[iop]{emulateapj}

\usepackage{times,graphicx}

\usepackage{float}
\usepackage{ifthen}
\usepackage{times}
\usepackage{natbib}
\usepackage{rotating}
\usepackage{units}
\newcommand {\apgt} {\ {\raise-.5ex\hbox{$\buildrel>\over\sim$}}\ }
\newcommand {\aplt} {\ {\raise-.5ex\hbox{$\buildrel<\over\sim$}}\ }

\makeatother \lefthead{{\sc Geach et al.}} \righthead{\sc A redline
starburst}

\begin{document}

\title{A redline starburst: CO(2--1) observations of an Eddington-limited
galaxy reveal star formation at its most extreme}

\author{J.\ E.\ Geach$^1$, R. C.\ Hickox$^2$, A. M.\ Diamond-Stanic$^3$, M. Krips$^4$, J. Moustakas$^5$,\\ C.\ A.\ Tremonti$^6$, A.\ L.\ Coil$^3$, P.\ H.\ Sell$^6$, G.\ H.\ Rudnick$^7$}

\altaffiltext{1}{Department of Physics, McGill University, Montr\'eal, Qu\'ebec, Canada.
jimgeach@physics.mcgill.ca}

\altaffiltext{2}{Department of Physics and Astronomy, Dartmouth College, Hanover, NH 03755, USA}
 
\altaffiltext{3}{Center for Astrophysics and Space Sciences, University of California, San Diego, La Jolla, CA 92093, USA} 

\altaffiltext{4}{Institut de Radioastronomie Millim\'etrique, 300 rue de la
Piscine F-38406 Saint Martin d'H\`eres, France}

\altaffiltext{5}{Department of Physics and Astronomy, Siena College, 515
Loudon Road, Loudonville, NY 12211}

\altaffiltext{6}{Department of Astronomy, University of Wisconsin-Madison, Madison, WI 53706, USA}

\altaffiltext{7}{Department of Physics and Astronomy, University of Kansas, Lawrence, KS 66045, USA}

\label{firstpage}

\begin{abstract}We report observations of the CO(2--1) emission of
SDSS\,J1506+54, a compact ($r_{\rm e}\approx135$\,pc) starburst galaxy at
$z=0.6$. SDSS\,J1506+54 appears to be forming stars close to the limit allowed
by stellar radiation pressure feedback models: the measured $L_{\rm
IR}/L'_{\rm CO}\approx1500$ is one of the highest measured for any galaxy.
With its compact optical morphology but extended low surface brightness
envelope, post-starburst spectral features, high infrared luminosity ($L_{\rm
IR}>10^{12.5}L_\odot$), low gas fraction ($M_{\rm H_2}/M_\star\approx15$\%),
and short gas depletion time (tens of\,Myr), we speculate that this is a
feedback-limited central starburst episode at the conclusion of a major
merger. Taken as such, SDSS\,J1504+54 epitomizes the brief closing stage of a
classic model of galaxy growth: we are witnessing a key component of spheroid
formation during what we term a `redline' starburst. \end{abstract}


\section{Introduction}

Theoretical studies suggest that the majority of the stellar mass in the
central regions of massive spheroids forms {\em in situ} in powerful, compact
starbursts (e.g.\ Narayanan et al.\ 2010). In these systems, radiative
feedback from stars is expected to produce an upper limit on the star
formation rate density ($\dot\Sigma_\star$) by limiting the density of the
star-forming gas ($\Sigma_{\rm gas}$, Thompson, Quataert \& Murray\ 2005;
Murray, Quataert \& Thompson 2005). One explanation for the small range in
central stellar surface densities observed in the cores of local spheroids is
that they formed in compact starbursts where the bulk of the gas reservoir
formed stars at the Eddington limit (Hopkins et al.\ 2010), however
observations have only just begun to probe the most extreme
$\dot\Sigma_\star$--$\Sigma_{\rm gas}$ in distant powerful starbursts that are
the antecedents of massive galaxies today.

Diamond-Stanic et al.\ (2012, DS12) reported the discovery, with the {\it
Wide-field Infrared Survey Explorer} ({\it WISE}), of a population of
moderate-redshift ($z\approx0.5$) galaxies that appear to be among the highest
density starbursts yet observed. {\em Hubble Space Telescope} ({\it HST})
imaging reveals that these systems are remarkably compact in the optical
bands, with $r_e$$\sim$100\,pc (DS12, P.\ H.\ Sell et al.\ 2013 in
preparation), while the optical spectra and UV--IR spectral energy
distributions imply high star formation rates (SFRs) approaching
10$^3$\,$M_\odot$\,yr$^{-1}$ (and thus very high $\dot\Sigma_\star$), with
little or no contribution from an active galactic nucleus (AGN). The galaxies
also exhibit extremely blueshifted (1000\,km\,s$^{-1}$) Mg\,{\sc
ii}\,$\lambda\lambda$2796, 2803 interstellar absorption lines (Tremonti et
al.\ 2007) indicative of powerful feedback.

In order to test if these compact galaxies represent the formation of a
massive stellar bulge via a feedback limited starburst, we require a
measurement of the cold gas reservoir, since Eddington limited star formation
places a theoretical upper limit on $\dot\Sigma_\star/\Sigma_{\rm gas}$ for
the actively star-forming gas. In this Letter we present new observations at
2\,mm of the most extreme galaxy in the DS12 sample with the Institut de
Radioastronomie Millim\'etrique (IRAM) Plateau de Bure Interferometer (PdBI).
We measure the CO(2--1) molecular emission line and derive the gas properties
of this remarkable system to test the picture of Eddington limited star
formation in a compact starburst galaxy.

\begin{figure*}
\centerline{\includegraphics[width=0.49\textwidth,angle=0]{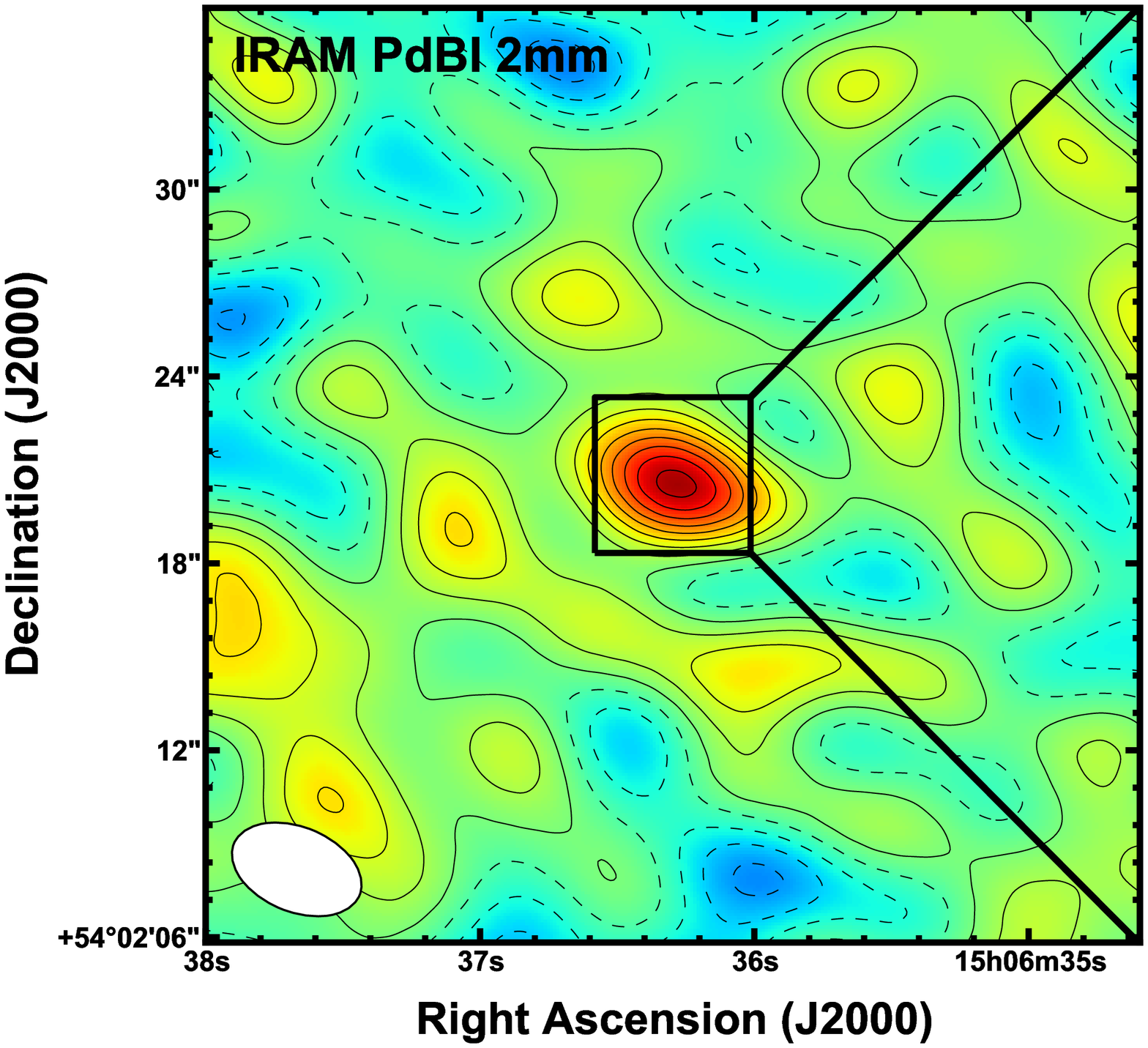}\includegraphics[width=0.49\textwidth,angle=0]{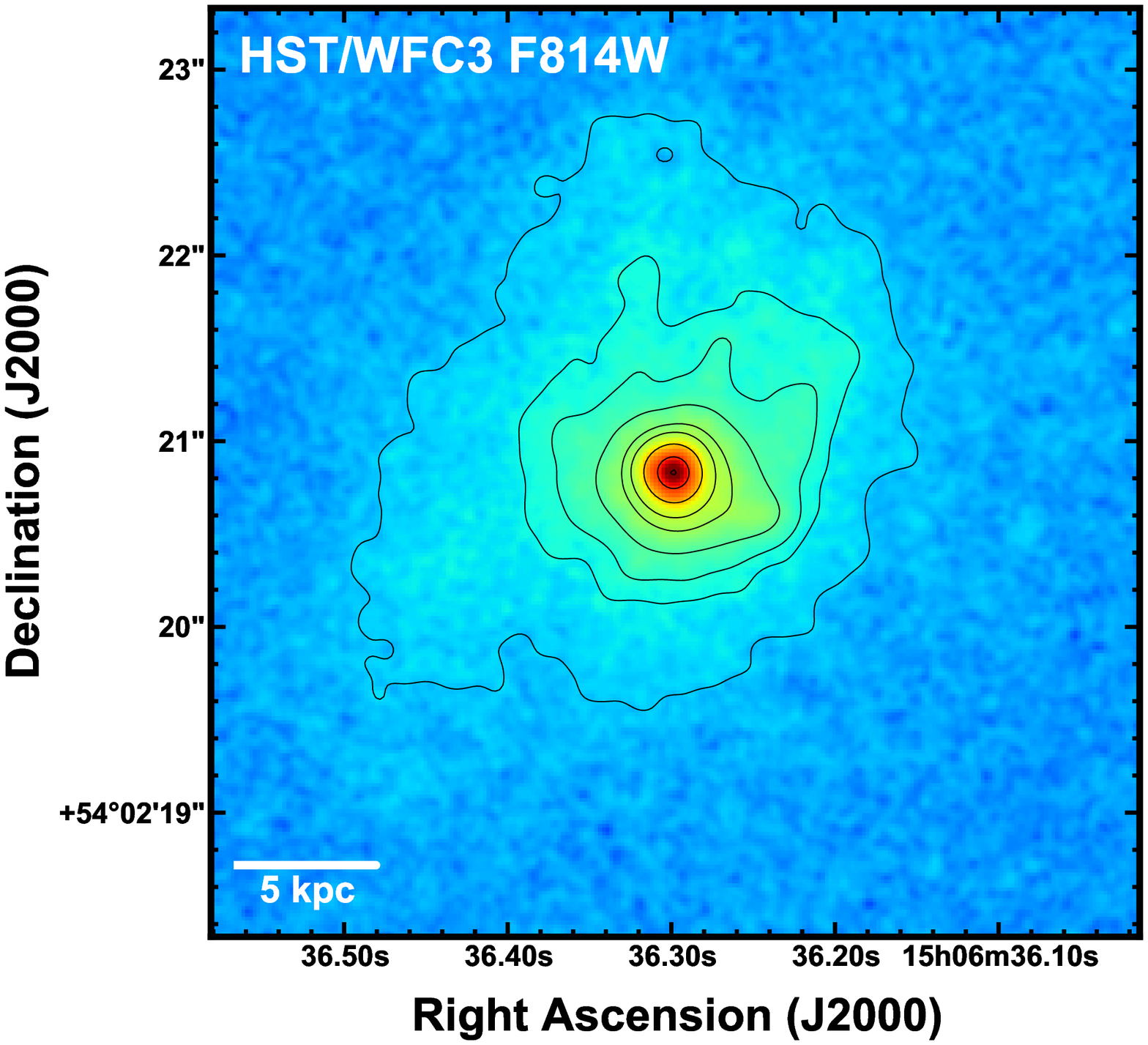}}
\caption{(left) Velocity integrated ($\Delta V=360$\,km\,s$^{-1}$) IRAM map of
the CO\,(2-1) emission of SDSS\,J1506+54. The white ellipse shows the
$4.4''\times2.7''$ beam, and contours are at levels of 0.33\,mJy (equivalent
to 1$\sigma$; negative values are shown as dashed contours), (right) Zoom-in
of $5''\times5''$ box (left) showing the {\it HST}/WFC3 F814W (rest-frame {\it
V}-band) image. The light is dominated by the compact core, $r_e=135$\,pc
(DS12), but fainter, diffuse optical emission is also visible over scales of
several kpc indicating a low surface brightness (LSB) envelope of UV emission
(contours are shown as a guide). More extensive LSB emission beyond this field
of view is also evident, at a level of $\mu_{\rm F814W}\approx25$\,
mag\,arcsec$^{-2}$, including a potential tidal tail feature (P. H. Sell et
al. in prep). This points to a merger origin for SDSS\,J1506+54 (\S6).}
\end{figure*}

\section{SDSS\,J150636.30+540220.9}

SDSS\,J150636.30+540220.9 at $z=0.608$ (hereafter `SDSS\,J1506+54') was
targeted because it is the most extreme system in the DS12 sample, in terms of
its star formation rate density. The effective radius measured in {\it HST}
WFC3 F814W (rest-frame {\it V}-band) imaging is $r_e\approx135$\,pc (Figure\
1), and even the most conservative estimates for the SFR result in densities
of $\dot\Sigma_{\star}\approx3000$\,$M_\odot$\,yr$^{-1}$\,kpc$^{-2}$. Fitting
to the {\it WISE} 12 and 22$\mu$m photometry, the integrated 8--1000$\mu$m
luminosity of SDSS\,J1506+54 is estimated to be in the range $\log(L_{\rm
IR}/L_\odot)=12.6$--$13.1$, assuming a representative range of spectral energy
distributions appropriate for star-forming galaxies (Chary \& Elbaz\ 2001,
Dale \& Helou\ 2002, Rieke et al.\ 2009). The dispersion in rest-frame
24$\mu$m luminosity using this range of templates is just 0.1\,dex. A
complementary estimate of $L_{\rm IR}$ is therefore made using $L_{24}$ using
the empirical correlation between $L_{24}$--$L_{\rm 8-1000}$ found by Rieke
et al.\ (2008); we find $\log(L_{\rm IR}/L_\odot)=12.78\pm0.06$, consistent with the estimate above. The SFR of SDSS\,J1506+54 is estimated to be in the range 340--
1400\,$M_\odot$\,yr$^{-1}$ assuming the calibration of Kennicutt\ (1998),
scaled to a Chabrier initial mass function. This is in good agreement with the
SFR estimated from stellar population fits to the UV-optical-IR photometry
(DS12).

The stellar mass is estimated from stellar population template fits to the
0.1--3$\mu$m photometry, with $\log_{10}(M_\star/M_\odot)=11.12\pm0.07$
(assuming a 0.1--100\,$M_\odot$ Chabrier initial mass function, see Moustakas
et al.\ 2013). Finally, the outflow velocity of the galactic wind is traced by
the interstellar medium Mg\,{\sc ii}\,$\lambda\lambda$2796, 2803 absorption
lines, which indicate a maximum $v_{\rm abs}\approx-1210$\,km\,s$^{-1}$
(Tremonti et al.\ 2007, DS12).

\subsection{Comments on possible AGN contribution}

The source does not appear to be dominated by an AGN. Although SDSS\,J1506+54
has the most luminous [O\,{\sc iii}]$\lambda$5007 and [Ne\,{\sc
{v}}]$\lambda$3426 emission of the DS12 sample ($\log(L_{\rm [O\,{III}]}/{\rm
erg\,s^{-1}})=42.1$ and $\log(L_{\rm [Ne\,{V}]}/{\rm erg\,s^{-1}})=41.1$),
only 4 counts were detected with {\em Chandra} in the 2--10\,keV X-ray band,
corresponding to a luminosity of
$L_X=2.9^{+4.4}_{-2.9}\times10^{42}$\,erg\,s$^{-1}$. Assuming typical ratios
between X-ray, [O\,{\sc iii}]$\lambda$5007 and [Ne\,{\sc {v}}]$\lambda$3426
luminosities, and accounting for possible small $\sim$30\% dust attenuation of
the narrow-line region based on the reddening derived from our fit to the SED
(DS12), the {\em Chandra} limits imply an X-ray absorption factor of
$\approx$10--20 for a buried AGN (DS12, Heckman et al.\ 2005, Gilli et al.\
2010). However, as DS12 argue, even if the X-ray attenuation factor was as
high as 100, the AGN would contribute less than half of the observed 22$\mu$m
luminosity (Diamond-Stanic et al.\ 2009, Gandhi et al.\ 2009). Thus, star
formation is likely to be the dominant power source of the bolometric emission
of SDSS\,J1506+54.

\section{IRAM Plateau de Bure 2\,mm Observations}

Observations were conducted with IRAM PdBI over June--July 2012 in
configuration D, using 5 antennas and the WideX correlator. We targeted the
CO(2--1) 230.54\,GHz rotational transition at $\nu_{\rm obs}=143.37$\,GHz
(2\,mm band) in the direction of SDSS\,J1506+5402. The total (on source)
integration time was 6.2\,hours (after flagging; scans on-source were
discarded for which the phases deviated by more than 45 degrees from the
solution), during which time the system temperature ranged between $T_{\rm
sys}=100$--$400$\,K ($\langle T_{\rm sys}\rangle\approx 200$--$250$\,K) for
precipitable water vapour in the range ${\rm pwv}=4$--$12$\,mm. Bandpass
calibrators were the sources 3C84, 2200+420 or 3C273, gain calibration was
performed with the sources 1418+546 and J1604+572 and the source MWC349 was
used for flux calibration. The flux calibration accuracy is $\sim$5--10\% at
2\,mm. Data were calibrated, mapped and analyzed using {\sc gildas}
(Guilloteau \& Lucas\ 2000).

\section{Results}

Figure\ 1 shows the velocity integrated CO(2-1) map of SDSS\,J1506+54 and
Figure\ 2 presents the 2\,mm spectrum around $\nu_{\rm obs}=143.37$\,GHz. The
CO(2--1) line is detected with high confidence at $>$8$\sigma$ in the
integrated map. The integrated line flux is $S\Delta V =
0.97\pm0.19$\,Jy\,km\,s$^{-1}$, corresponding to a luminosity of $L'_{\rm
CO(2-1)}=(4.8\pm0.9)\times10^{9}$\,K\,km\,s$^{-1}$\,pc$^2$. Uncertainties are
conservative; they are estimated via a bootstrap analysis, where we
generate a series of realizations of the spectrum, adding noise to each
40\,km\,s$^{-1}$ channel, selected from a Gaussian distribution with
$\sigma=1.1$\,mJy. The standard deviation of the ensemble of integrated
spectra are taken to be the 1$\sigma$ error of the measured line flux.

The line profile is adequately fit by a single Gaussian profile
($\chi^2/\nu=1.1$), with a velocity width ${\rm FWHM}=283\pm
31$\,km\,s$^{-1}$ (Fig.\ 2). Assuming the average inclination angle
$i=30^\circ$, the true width could be as large as $\sim$560\,km\,s$^{-1}$.
Galaxy-integrated CO spectra often exhibit double peaked line profiles,
indicative of a rotating disc or ring (e.g.\ Downes \& Solomon\ 1998). There
are some hints of this in the current spectrum, and so we also fit a double-Gaussian profile (fixing the individual line peaks
and individual dispersions to be equal). The fit is formally $\chi^2/\nu=0.9$,
not significantly improved over the single component fit. The peak-to-peak
velocity is $\Delta V=(147\pm12)$\,km\,s$^{-1}$, and the line dispersion,
which includes the 1-dimensional turbulent velocity dispersion of the gas, is
$\sigma=(49\pm6)$\,km\,s$^{-1}$.

At first glance, it seems puzzling that the line width is not even
larger, if the mass distribution is as compact as suggested by the
rest-frame UV morphology. There are two main possibilities: (i) the $r_e$
for the stellar mass is more widely distributed than $r_e$ of light
measured in the F814W band (near-infrared observations at the same
resolution will be required to test this) such that the environment
dominated by star formation is compact ($r\approx r_e$), but molecular
gas dominated, or (ii) the CO-emitting gas is more widely distributed
than the compact optical emission. The lack of size
and orientation constraints for the CO(2--1) emission makes any estimate of
$M_{\rm dyn}$ uncertain, however it would be unusual if the bulk of the
gas reservoir (and presumably the concomitant dust) was beyond $r_e$ and
configured in a way that did not completely obscure the UV-bright core.

\section{Interpretation and discussion}

\begin{figure}
\centerline{\includegraphics[width=0.75\linewidth,angle=-90]{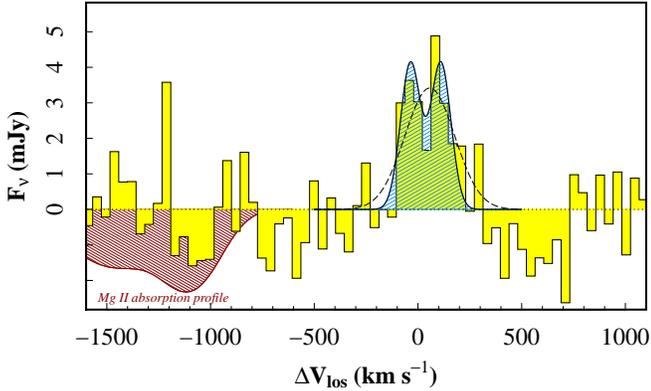}}
\caption{The 2\,mm spectrum of SDSS\,J1506+54 obtained with IRAM PdBI, binned
to a resolution of 40\,km\,s$^{-1}$. The channel noise (r.m.s.) is 1.1\,mJy
per 40\,km\,s$^{-1}$ channel. We consider two model fits to the data: a single
Gaussian profile, and a double Gaussian (\S4). The latter is an approximation
to the line profile of a rotating disc or ring, and the observed spectrum does
show hints of a double-peaked profile. We note that there is a 3.2$\sigma$
spike in the spectrum close to the outflow velocity (as traced by blue-shifted
Mg\,{\sc ii} absorption; we show the best-fit absorption profile of the
Mg\,{\sc ii} doublet as a shaded region on an arbitrary flux scale). We
acknowledge this is currently far too tentative to draw any conclusions, but
we remark that -- if real -- this could be indicative of cold gas entrained in
the outflow (\S5.4).} \end{figure}

\subsection{Comparison with other galaxies}

The measured $L_{\rm IR}/L'_{\rm CO}\approx1500$ is one of the highest
measured for any star-forming galaxy (Fig.\ 3). Even if we
accounted for an AGN contribution of 25\% to our lowest estimate of $L_{\rm
IR}$, SDSS\,J1506+54 is at the extreme tail of the $L_{\rm IR}/L'_{\rm CO}$
distribution, with $L_{\rm IR}/L'_{\rm CO}\approx550$ in this case. There are
a handful of other star-forming galaxies with similar $L_{\rm
IR}/L'_{\rm CO}$ properties; for example, the $z=0.575$ hyper-LIRG
IRAS\,F00235+1024 (Combes et al.\ 2011) and the $z=0.633$ ULIRG
IRAS\,F10398+3247 (Combes et al.\ 2012) both have CO line luminosities
measured in the 2--1 transition, and are also at the extreme end of the
$L_{\rm IR}/L'_{\rm CO}$ distribution (Fig.\ 3). Several galaxies in the
sample of Combes et al.\ (2012) have $L'_{\rm CO}$ upper limits implying
extreme $L_{\rm IR}/L'_{\rm CO}$ and of these, two are obviously hosting AGN,
as evident from their optical spectra.

Like SDSS\,J1506+54, F00235+1024 is not dominated by an AGN, although there is thought to be a non-negligible ($\approx$30\%) AGN contribution to
its total infrared emission (Verma et al.\ 2002, Farrah et al.\ 2002).
F10398+3247 could also contain a deeply buried AGN, as it has a deep silicate
absorption feature ($\tau_{\rm Si}>3$, Dartois \& Mu\~noz-Caro\ 2007), often
associated with nuclei blanketed by a dense screen of carbonaceous and
silicate grains. The key difference between these galaxies and
SDSS\,J1506+54 is in optical morphology: F00235+1024 and F10398+3247 are interacting systems, where multiple UV/optical clumps can be
identified in {\it HST} optical and NIR imaging (Stanford et al.\ 2000, Farrah
et al.\ 2002, Combes et al.\ 2012), thus it is unclear whether these galaxies
support such extreme $\dot\Sigma_\star$.

\begin{figure}
\centerline{\includegraphics[width=\linewidth,angle=-90]{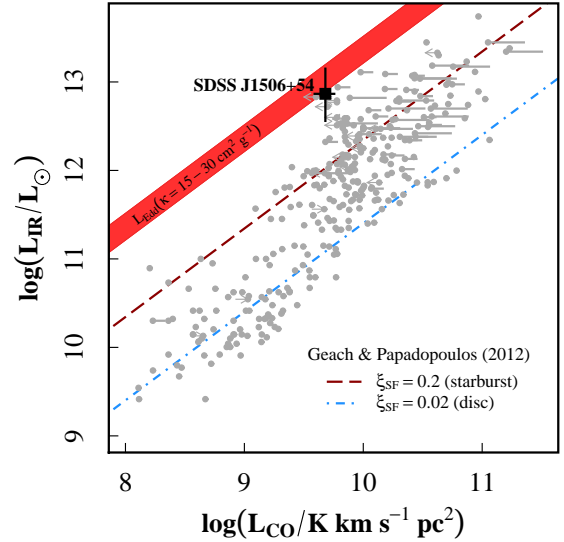}}
\caption{$L_{\rm IR}$--$L'_{\rm CO}$ $(J_{\rm up}\leq3)$ plane for
star-forming galaxies, showing the compilation of data for super
star clusters, normal discs, local starburst galaxies, ULIRGs,
submillimeter galaxies, hyper-LIRGs and circumnuclear starbursts from Andrews
\& Thompson\ (2011). We plot $L'_{\rm CO}$ $J_{\rm up}>1$ uncorrected for
sub-thermal excitation, but horizontal lines indicate how the points would
move for corrections of $0.8\leq r_{\rm 21}\leq1$ and $0.5\leq r_{\rm
31}\leq1$ respectively. The shaded region shows $L_{\rm IR}\approx L_{\rm
Edd}$ for optically thick dust emission (see \S5.2). We also show predictions
from Geach \& Papadopoulos\ (2012) which models the global gas reservoir as a combination of
dense star-forming gas and a diffuse quiescent cold gas phase. SDSS\,J1506+54
is close to the Eddington limit, with only a handful of other galaxies exhibiting similar ratios (Combes et al.\ 2011 \& 2012).} \end{figure}

\subsection{Eddington-limited star formation}

How do we interpret such extreme systems? Can $L_{\rm IR}/L'_{\rm CO}$ be
driven to such high values through star formation alone? In individual
galaxies, isolated star-forming cores might be Eddington limited, which is to
say $\dot\Sigma_\star$ is capped by radiation pressure from the recently
formed O and B associations (e.g.\ Scoville\ 2004, Shirley et al.\ 2003).
However, when taken as a whole, galaxies form stars at sub-Eddington rates due
to intermittency -- the fact that only a small fraction of the cold gas
reservoir is actually forming stars (Downes \& Solomon\ 1998, Murray, Quataert
\& Thompson\ 2005, Andrews \& Thompson\ 2011). Under certain conditions
however, one could envision a scenario where the majority of the gas reservoir
could be driven to form stars at the Eddington limit.
SDSS\,J1506+54 is an excellent candidate for this phenomenon.

In the absence of nuclear heating, and assuming the model of momentum-driven
feedback (e.g.\ Murray, Quataert
\& Thompson\ 2005) the upper limit of $L_{\rm IR}/L'_{\rm CO}$ for
star-forming galaxies is set by the Eddington luminosity. For optically thick
dust emission ($\tau_{\rm 100\mu m}>1$):

\begin{equation}
	L_{\rm Edd} = \frac{4\pi G c}{\kappa}X_{\rm CO}L'_{\rm CO}
\end{equation}

\noindent (Andrews \& Thompson\ 2011), where here we set $X_{\rm CO}=3$ (see
\S5.3) and $\kappa$ is the Rosseland-mean dust opacity. In Figure\ 3 we show
the Eddington luminosity predicted for $\kappa = 5$--$10$\,${\rm cm}^2\,{\rm
g}^{-1}$\,$(f_{\rm dg}/f^{\rm MW}_{\rm dg})$, where $f^{\rm MW}_{\rm dg} =
\nicefrac{1}{150}$ is the Milky Way dust-to-gas mass ratio (see Andrews \&
Thompson\ 2011). We assume a dust-to-gas ratio of $f_{\rm
dg}\approx\nicefrac{1}{50}$, appropriate for dusty star-forming galaxies (e.g.
Kov\'acs et al.\ 2006). Increasing either $f_{\rm dg}$ (as might be
appropriate in highly obscured galaxies), or $\kappa$, or decreasing $X_{\rm
CO}$, lowers $L_{\rm Edd}$ for a fixed $L'_{\rm CO}$.

Our chosen values aim to be conservative in this respect, such that no
star-forming galaxies can be described as super-Eddington, but we caution the
reader that in this parameter space the Eddington limit is not expected to be
a hard edge.  A simple framework
that explains the {\it intrinsic} range in $L_{\rm IR}/L'_{\rm CO}$ is the
model of Geach \& Papadopoulos\ (2012) that predicts the minimal dense gas
mass for a given $L_{\rm IR}$, and a range of gas mass ratios $\xi_{\rm
SF}=M_{\rm dense}/M_{\rm total}$. Typically, $\xi_{\rm SF}\approx0.02$ for
quiescent discs and $\xi_{\rm SF}\approx0.2$ for ULIRG-like systems (Fig.\ 3).
The Eddington limit in equation\ (1) is approached as $\xi_{\rm
SF}\rightarrow1$, indicating that the majority of the gas in SDSS\,J1506+54 is
in a dense, active phase and the intermittency is close to unity.

\subsection{Estimating the gas mass}

Recent studies of the local (U)LIRG population where well-sampled CO and
$^{13}$CO ladders are available (Papadopoulos et al.\ 2012a) show that, when
high-{\it J} CO or heavy rotor (HCN, CS) line transitions are included in
radiative transfer models, the $M_{\rm H_2}/L'_{\rm CO}=X_{\rm CO}$ factor
increases by a factor $\sim$3--10 compared to the widely applied $X_{\rm
CO}=0.8$ conversion\footnote{we omit the units of $M_\odot({\rm
K\,km\,s^{-1}\,pc^2})^{-1}$} usually adopted for ULIRG-like systems
(Papadopoulos et al.\ 2012b). The explanation is that in the supersonic
turbulent discs of ULIRGs, a significant fraction of the gas mass is expected
to be in a dense ($n>10^4$\,cm$^{-3}$) phase (Padoan \& Nordlund\ 2002,
Papadopoulos et al.\ 2012, Papadopoulos \& Geach\ 2012). This is not well
traced by single low-{\it J} CO tracers, despite the fact that these were
originally used to calibrate the value of $M_{\rm H_2}/L'_{\rm CO}$ in ULIRGs
(Downes, Solomon \& Radford\ 1993, Solomon et al.\ 1997, Downes \& Solomon\
1998).

Here we assume that the majority of the gas reservoir is indeed in a dense,
turbulent phase. Setting $X_{\rm CO}=3$ and $r_{21}=0.8$ (Papadopoulos et al.\
2012a, Geach \& Papadopoulos\ 2012), we estimate $M_{\rm
H_2}=(1.9\pm0.3)\times10^{10}M_\odot$. This is in agreement with the {\rm
minimum} dense gas mass expected if $L_{\rm IR}$ is powered by
Eddington-limited star formation, since one expects a maximum $\epsilon_\star=
L_{\rm IR}/M_{\rm dense}\approx500\,(L_\odot/M_\odot)$ (Scoville 2004,
Thompson et al.\ 2005, Thompson\ 2009). The $\epsilon_\star$ expected in
feedback models is indeed close to the value actually measured both for
resolved star-forming cores in spiral discs and even entire starbursts
(Scoville\ 2004, Shirley et al.\ 2003). Similar values of $\epsilon_\star$ are
obtained when one considers the mass-to-light ratio of a deeply
dust-enshrouded ($\tau_{\rm IR}>1$) zero-age main sequence (ZAMS) population,
with $L_{\rm ZAMS}\approx L_{\rm IR}$ (Downes \& Solomon\ 1998). It is
important to note that, regardless of the exact gas mass calibration, assuming
that the majority of the infrared emission originates from enshrouded
star-forming regions, the empirical scalings in Figure\ 3 clearly indicates
that SDSS\,J1506+54 is close to the Eddington limit.

\subsection{Hints of gas entrained in a wind}

The compact starburst in SDSS\,J1506+54 should be effective in the launching
of a powerful momentum-driven galactic wind, powered by radiation pressure on
dust grains intermingled with the ISM, and the ram-pressure of supernovae
detonations (Murray, Quataert \& Thompson\ 2005). The strongly blue-shifted
Mg\,{\sc ii} lines are excellent evidence that this is the case (Tremonti et
al.\ 2007, DS12).

The 2\,mm spectrum of SDSS\,J1506+54 exhibits a marginally significant
(3.2$\sigma$) narrow ($\sigma<40$\,km\,s$^{-1}$) spike close to the outflow
velocity traced by Mg\,{\sc ii} (no RFI or telluric contaminant is expected at
this frequency). We can draw no conclusions on such a low-significance
feature, but we remark that if cold gas has become swept-up in the outflow
traced by Mg\,{\sc ii} absorption, and have survived
shock-heating, then we might expect to detect entrained cold gas in this way.

\section{SDSS\,J1506+54 as an Eddington-limited starburst induced in a
late-stage merger}

\subsection{Formation}

We speculate that SDSS\,J1506+54 is the remnant of a major merger. In this picture angular momentum losses caused by the
dissipation of energy in shocks and gravitational torques cause the collapse
of disc-gas to the central regions in each parent, and subsequently into a
single merged core. This is a classic picture of galaxy growth that has been
supported for many years by both theory, numerical simulations and observation
(Toomre \& Toomre\ 1972, Sanders et al.\ 1988, Hernquist\ 1989, Barnes \&
Hernquist\ 1991, Sanders \& Mirabel\ 1996, Schweizer et al.\ 1996, Mihos \&
Hernquist\ 1996, Wuyts et al.\ 2010, Hopkins et al.\ 2012).

Tidal interactions in earlier stages of the merger might well be responsible
for the faint extended (over several kpc) optical light (Figure\ 1, P. H. Sell
et al.\ in prep). Episodes of massive star formation must have occurred in the
recent past (0.5--1\,Gyr), with the A stars formed in (or moved to) relatively
unobscured regions, in order to produce strong H$\delta$ absorption observed
in the optical spectrum. Indeed, the presence of post-starburst spectral
features is entirely consistent with the typical timescales for equal-mass
mergers (Lotz et al.\ 2008). 

\subsection{Fate}

In the absence of additional cooling of gas, SDSS\,J1506+54 must be close to
the cessation of star formation. At the current rate the gas will be consumed
within a few tens of Myr, increasing the stellar mass by just $\sim$15\%. This
will place the galaxy in the exponential tail of the stellar mass function at
$z\approx0.5$ (Moustakas et al.\ 2013), but the current starburst contributes
only a modest fraction of the total mass. Nevertheless, this small addition can have
important consequences for the properties of the descendant (Robaina et al.\
2009, Hopkins et al.\ 2010, 2012, Hopkins \& Hernquist\ 2010).

Signatures of compact Eddington-limited starburst episodes could be found in
the `fossil record' of local spheroids. Integral field observations of the
cores of local elliptical galaxies reveal examples of compact ($\sim$100\,pc),
`kinematically decoupled', relatively young ($\aplt$5\,Gyr) stellar components
in a subset of the population (e.g.\ McDermid et al.\ 2006). The origin of
these structures could well be found in powerful, compact starbursts such as
the one we present. Star formation driven at extremely high $\dot\Sigma_\star$
could also have a critical impact on the form of the stellar initial mass
function, due to the corresponding high cosmic ray densities in such
environments (Papadopoulos et al.\ 2011).

\section{Summary}

We have presented observational evidence of what we term a `redline'
starburst: a galaxy forming stars close to the Eddington limit assuming
current models of stellar feedback. The measured $L_{\rm IR}/L'_{\rm
CO}\sim1500$ is one of the highest measured for any galaxy. Our conclusion is that the SDSS\,J1504+54 system is a final stage merger,
undergoing a high intensity circumnuclear starburst. If we
are witnessing the intense -- but fleeting -- closing stages of the assembly
of a massive stellar bulge, SDSS\,J1506+54 represents a unique opportunity to
study the physics associated with this important evolutionary phase, and the
mechanics of star formation at its most extreme.

\section*{Acknowledgements} 

We thank the anonymous referee for a constructive report that improved this
paper. We are also grateful to Brett Andrews, Padelis Papadopoulos, Nathan
Bastian and Aday Robaina for helpful discussions and advice. J.E.G. is
supported by a Banting Fellowship. A.M.D. acknowledges support from the
Southern California Center for Galaxy Evolution. P.H.S. acknowledges support
through the NASA grant HST-GO-12019. This work was also supported by {\it
Chandra} grant \#tGO0-11135.

\label{lastpage}
\end{document}